\documentclass[12pt,a4paper,dvips]{article}
\usepackage{a4p}
\usepackage{times}
\usepackage{cite}
\usepackage{mcite}
\usepackage{amssymb}
\usepackage{graphicx}
\usepackage{l3_note}
\usepackage{ifthen}

\date{November 16, 2001}
\revised{January 22, 2001}
\l3note{2722}

\newlength{\capwidth}
\begin{document}
\flushbottom
\begin{titlepage}
%
\title{On the Experimental Effects of the Off-shell Structure in
        Anomalous Neutral Triple Gauge Vertices}
\begin{Authlist}
J.~Alcaraz\\
CERN / CIEMAT
\end{Authlist}
\begin{abstract}
   We discuss differences between on-shell and off-shell treatments
in the search for anomalous neutral triple gauge couplings in $\mathrm{e^+ e^-}$ collisions.
We find that the usual on-shell framework represents an optimal starting
point, covering all scenarios in which a reasonable experimental 
sensitivity is expected.
We show that off-shell effects lead to 
negligible deviations at the experimental level, provided that 
$\mathrm{e^+ e^-}\rightarrow\mathrm{f}\bar{\mathrm{f}}\gamma$ and $\mathrm{e^+ e^-}\rightarrow\mathrm{f}\bar{\mathrm{f}}\mathrm{f^\prime}\mathrm{\overline{f}^\prime}$ analyses are 
performed in regions where $\mathrm{Z}^*\rightarrow\mathrm{f}\bar{\mathrm{f}},\mathrm{f^\prime}\mathrm{\overline{f}^\prime}$ 
production is dominant. For consistency reasons, we advocate the use 
of a natural extension of the on-shell definitions, which takes 
into account the correct off-shell dependences. Contrary to what has been 
recently suggested in the literature, we find that no 
$SU(2)_L\times U(1)_Y$ constraints among neutral triple gauge couplings can 
be imposed in a general case.
\end{abstract}

\end{titlepage}

\section{Introduction}

\indent

  The measurement of triple gauge boson couplings is one of the 
main items in the physics program
of present and future colliders \cite{tgc}. In this context, 
anomalous neutral triple gauge couplings (NTGC), which are not 
present in the Standard Model (SM) at tree level, constitute an interesting
possibility for New Physics \cite{gounaris_np}. Tevatron \cite{cdf,d0} 
and LEP collider experiments \cite{aleph,delphi,l3,opal} have carried out 
systematic searches for $\mathrm{Z}VV$ couplings, where $V$ 
denotes any of the two SM neutral gauge bosons ($\mathrm{Z}$ or $\gamma$). 

  Recently it has been claimed \cite{gounaris_offshell,rakshit} that off-shell 
effects in anomalous couplings can not be ignored, and that the spectrum of 
possible coupling structures may be larger. LEP analyses on the search for 
anomalous off-shell couplings have followed \cite{delphi}.
  The aim of this paper is to clarify the situation in what 
respects the different NTGC sets and conventions, and
the implications of these choices on present experimental limits. 

  The study is organized as follows. The first section introduces the 
usual convention employed in the search for anomalous NTGCs. Next we 
present a general discussion on NTGCs arising at 
the lowest order in $\sqrt{s}/\Lambda$, where $\Lambda$ represents the scale of 
New Physics. A new convention for the NTGC structures will be suggested 
at this stage. It will be shown that the new 
convention should lead to no changes in what respects present 
experimental results \cite{cdf,d0,aleph,delphi,l3,opal}. A different 
approach will be used in order to build up the 
off-shell dependences for the remaining (higher order) NTGCs.
The study will be completed with a short discussion on 
the experimental consequences of imposing $SU(2)_L\times U(1)_Y$ SM 
symmetry constraints. The conclusions are presented in the last section.

\section{The standard convention: on-shell anomalous couplings}

\indent

\begin{figure}[htbp]
\begin{center}
    \includegraphics*[width=0.48\textwidth]{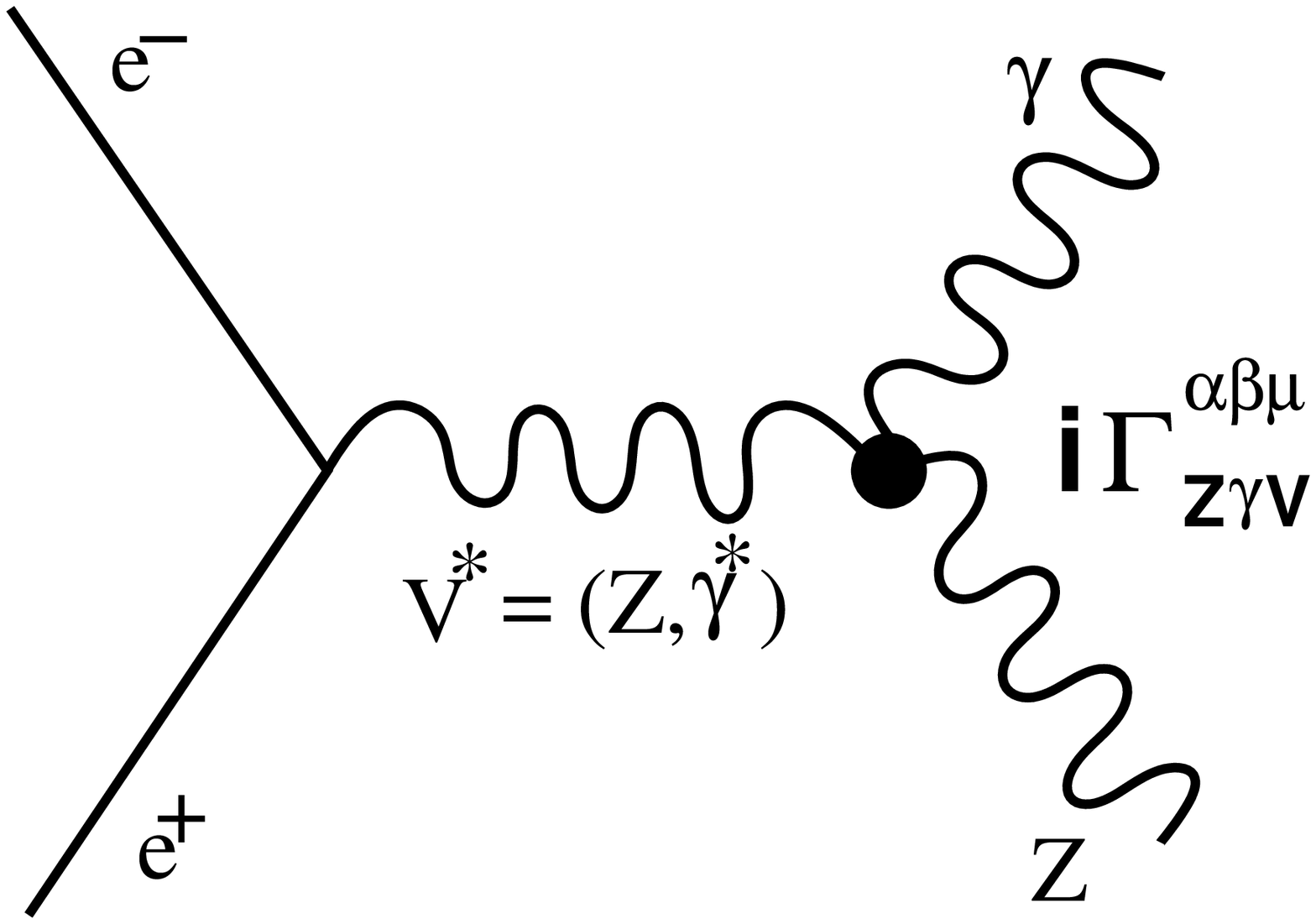}
    \includegraphics*[width=0.48\textwidth]{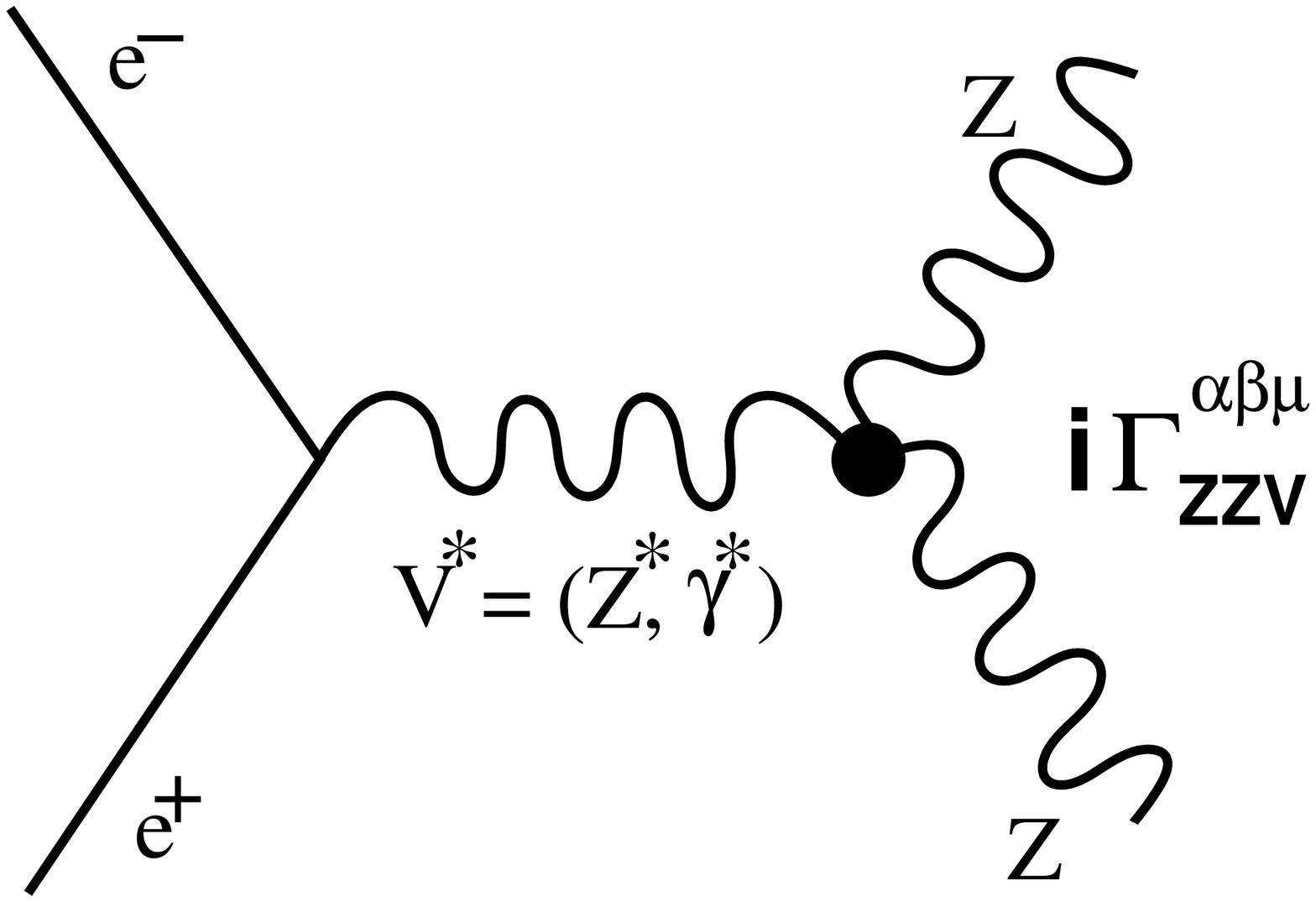}
\end{center}
\caption{ Anomalous vertex structures for $\mathrm{Z}\gamma V$ (left) and $\mathrm{Z}\mathrm{Z} V$ (right) 
anomalous couplings.}
\label{fig:anom}
\end{figure}

  The usual definition of anomalous NTGCs
is obtained from the vertex structures (see Figure \ref{fig:anom}):
\begin{eqnarray}
\Gamma^{\alpha\beta\mu}_{\mathrm{Z}\gamma V} = i~e\frac{q^2_V-m_V^2}{m_{\mathrm{Z}}^2} 
     \left\{ \right.
     h_1^V~( q_\gamma^\mu g^{\alpha\beta}-q_\gamma^\alpha g^{\beta\mu} )\nonumber\\
  ~+~h_2^V~\frac{ q_V^\alpha}{m_{\mathrm{Z}}^2}~
   (q_\gamma q_V~g^{\beta\mu}-q_\gamma^\mu~
   q_V^\beta ) \nonumber \\
  + h_3^V~\epsilon^{\alpha\beta\mu\rho}~q_{\gamma\rho}\nonumber\\
  ~+~h_4^V~\frac{q_V^\alpha}{m_{\mathrm{Z}}^2}~
   \epsilon^{\mu\beta\rho\sigma}~q_{V\rho}~
   q_{\gamma\sigma} \left. \right\}\label{eq:defzg}
\end{eqnarray}

\noindent
for the $\mathrm{e^+ e^-}\rightarrow\mathrm{Z}\gamma$ case, and:
\begin{eqnarray}
\Gamma^{\alpha\beta\mu}_{{\mathrm{Z}_1}{\mathrm{Z}_2} V} = i~e\frac{q^2_V-m_V^2}{m_{\mathrm{Z}}^2} 
      \left\{ \right.  f_4^V~( q_V^\alpha~g^{\beta\mu}
     + q_V^\beta~g^{\mu\alpha} ) \nonumber\\
   ~+~f_5^V~\epsilon^{\alpha\beta\mu\rho}~(q_{\mathrm{Z}_1\rho}-q_{\mathrm{Z}_2\rho})
   \left. \right\}\label{eq:defzz}
\end{eqnarray}

\noindent
for the $\mathrm{e^+ e^-}\rightarrow\mathrm{Z}\mathrm{Z}$ case. The momenta of the particles in the vertex are denoted 
by $q_V$ (ingoing) and $q_{\mathrm{Z}}$,$q_\gamma$,$q_{\mathrm{Z}_1}$,$q_{\mathrm{Z}_2}$ (outgoing). 
The electromagnetic coupling,  
$e=\sqrt{4\pi\alpha}$, and the $\mathrm{Z}$ mass, $m_{\mathrm{Z}}$, appear as arbitrary
constant factors.

  The anomalous $\mathrm{Z}\gamma V$ couplings $h_1^V,h_2^V$ ($V=\mathrm{Z},\gamma$) correspond 
to CP violating terms, whereas $h_3^V,h_4^V$ are related 
to CP conserving ones. The anomalous $\mathrm{Z}\mathrm{Z} V$ couplings 
$f_4^V$ lead to CP violating interactions, whereas $f_5^V$ are associated 
to a CP conserving structure. All terms violate charge conjugation.

  Both parametrizations were proposed for the first 
time in \cite{hagiwara}. For the $\mathrm{e^+ e^-}\rightarrow\mathrm{Z}\gamma$ case, the original 
proposal had to be modified \cite{gounaris_ifactor} 
(an extra $i$ factor was included) in order to work with Hermitian Lagrangians
for real values of the anomalous couplings. 

  The previous vertex expressions are the most general ones preserving Lorentz 
and electromagnetic gauge invariance, assuming that the bosons in the 
final state are on shell. Let us comment on some features related to the 
arbitrary factors in the convention:

\begin{itemize}
  \item The strength of the coupling is assumed to be electromagnetic,
        but it should be substituted in general by a coupling $g$, of order 
        one: 
                 $$e \rightarrow g~\sqrt{4\pi}$$

  \item $h_1^V,h_3^V,f_4^V,f_5^V$ are accompanied by a $m_{\mathrm{Z}}^{-2}$ factor. 
        They correspond to vertices arising from Lagrangians of dimension
        six or higher. It is convenient to reinterpret them in terms of 
        the new physics scale $\Lambda$:
             $$\frac{e}{m_{\mathrm{Z}}^2} \rightarrow \frac{g~\sqrt{4\pi}}{\Lambda^2}$$

  \item $h_2^V,h_4^V$ are accompanied by a $m_{\mathrm{Z}}^{-4}$ factor, and only appear
        via Lagrangian terms of dimension eight or higher. Similarly to the 
        previous case, the $m_{\mathrm{Z}}^{-4}$ dimensional factor could be substituted 
        by $\Lambda^{-4}$:
             $$\frac{e}{m_{\mathrm{Z}}^4} \rightarrow \frac{g~\sqrt{4\pi}}{\Lambda^4}$$
\end{itemize}

   Since this is just a matter of convention, adopted by all experiments 
until now, we are not proposing a redefinition in terms of scales of new 
physics. We just point out that if the sensitivities 
to $h_1^V,h_3^V$ and $h_2^V,h_4^V$ at center-of-mass energies $\sqrt{s}\gtrsim m_{\mathrm{Z}}$ 
turn out to be quite similar this is an artifact of the $m_{\mathrm{Z}}^2$ factors in 
the convention. The actual sensitivity to the New Physics scale $\Lambda$ is 
reduced in general for the higher dimension terms associated to $h_2^V,h_4^V$.

   In general, all these couplings behave as complex form-factors, with a 
dependence on $\sqrt{s}$. That is the case of the SM and of the 
Minimal Supersymmetric Standard Model (MSSM) close to the electroweak scale
\cite{gounaris_np,rakshit}. In the case of New Physics at a scale 
$\Lambda \gg \sqrt{s}$ the imaginary parts and $\sqrt{s}$ dependences can be 
ignored, since they are suppressed by powers of $(s/\Lambda^2)^n$.

\section{Off-shell couplings}

\indent

 At the lowest dimension (six), only the following operators contain 
sensible~\footnote{$\partial_\mu Z^\mu$ terms are ignored. They
are only relevant for off-shell decays into very massive fermions, like 
$\mathrm{Z}^*\rightarrow\mathrm{t}\bar{\mathrm{t}}$ \cite{hagiwara}.} 
$\mathrm{Z} V V$ vertex information:
\begin{eqnarray}
   {\mathcal O}_6^A & = & 
    \tilde{Z}_{\mu\nu} (\partial_\rho Z^{\rho\mu}) Z^\nu \\
   {\mathcal O}_6^B & = & 
    \tilde{F}_{\mu\nu} (\partial_\rho Z^{\rho\mu}) Z^\nu \\
   {\mathcal O}_6^C & = & 
    \tilde{Z}_{\mu\nu}  (\partial_\rho F^{\rho\mu}) Z^\nu \\
   {\mathcal O}_6^D & = & 
    \tilde{F}_{\mu\nu}  (\partial_\rho F^{\rho\mu}) Z^\nu \\
   \tilde{\mathcal O}_6^A & = & 
    {Z}_{\mu\nu} (\partial_\rho Z^{\rho\mu}) Z^\nu \\
   \tilde{\mathcal O}_6^B & = & 
    {F}_{\mu\nu} (\partial_\rho Z^{\rho\mu}) Z^\nu \\
   \tilde{\mathcal O}_6^C & = & 
    {Z}_{\mu\nu}  (\partial_\rho F^{\rho\mu}) Z^\nu \\
   \tilde{\mathcal O}_6^D & = & 
    {F}_{\mu\nu}  (\partial_\rho F^{\rho\mu}) Z^\nu
\end{eqnarray}

\noindent
where $F^{\mu\nu}$ and $Z^{\mu\nu}$ are the tensor fields associated to the 
photon and to the ${\mathrm Z}$ particle, respectively, and 
$\tilde{F}^{\mu\nu} \equiv \epsilon^{\mu\nu\rho\delta} F_{\rho\delta}$.
These terms give rise to anomalous vertices which we will parametrize as 
follows:
\begin{eqnarray}
  \Gamma^{\alpha\beta\mu}_{\mathrm{Z}\mathrm{Z}\mathrm{Z}}\rightarrow & ie~\frac{f_5^Z}{m_{\mathrm{Z}}^2} & \left[
   q^2_1 \epsilon^{\alpha\beta\mu\rho}~(q_{2\rho}-q_{3\rho})
 + q^2_2 \epsilon^{\alpha\beta\mu\rho}~(q_{3\rho}-q_{1\rho})
 + q^2_3 \epsilon^{\alpha\beta\mu\rho}~(q_{1\rho}-q_{2\rho})
                                      \right] \label{eq:firstoffshell} \\
  \Gamma^{\alpha\beta\mu}_{\mathrm{Z}\gamma\mathrm{Z}}\rightarrow & ie~\frac{h_3^Z}{m_{\mathrm{Z}}^2} & 
    \left[ (q^2_3-q^2_1)~\epsilon^{\alpha\beta\mu\rho}~q_{2\rho}  \right] \\
  \Gamma^{\alpha\beta\mu}_{\mathrm{Z}\mathrm{Z}\gamma}\rightarrow & ie~\frac{f_5^\gamma}{m_{\mathrm{Z}}^2} & \left[
   q^2_3~\epsilon^{\alpha\beta\mu\rho}~(q_{1\rho}-q_{2\rho}) \right]\\
  \Gamma^{\alpha\beta\mu}_{\mathrm{Z}\gamma\gamma}\rightarrow & ie~\frac{h_3^\gamma}{m_{\mathrm{Z}}^2} & \left[
    q^2_3 \epsilon^{\alpha\beta\mu\rho}~q_{2\rho}
  - q^2_2 \epsilon^{\alpha\beta\mu\rho}~q_{3\rho} \right] \\
  \Gamma^{\alpha\beta\mu}_{\mathrm{Z}\mathrm{Z}\mathrm{Z}}\rightarrow & ie~\frac{f_4^Z}{m_{\mathrm{Z}}^2} & \left[
 -q^2_1~(q_1^\beta  g^{\mu\alpha}+q_1^\mu   g^{\alpha\beta})
 -q^2_2~(q_2^\alpha g^{\beta\mu} +q_2^\mu   g^{\alpha\beta})
 -q^2_3~(q_3^\alpha g^{\beta\mu} +q_3^\beta g^{\mu\alpha})
          \right] \\
  \Gamma^{\alpha\beta\mu}_{\mathrm{Z}\gamma\mathrm{Z}}\rightarrow & ie~\frac{h_1^Z}{m_{\mathrm{Z}}^2} & \left[
      (q^2_3-q^2_1)~(
   g^{\alpha\beta}q_2^\mu - g^{\beta\mu}q_2^\alpha ) \right]\\
  \Gamma^{\alpha\beta\mu}_{\mathrm{Z}\mathrm{Z}\gamma}\rightarrow & ie~\frac{f_4^\gamma}{m_{\mathrm{Z}}^2} & \left[
        -q^2_3~( 
   g^{\beta\mu}q_3^\alpha + g^{\mu\alpha}q_3^\beta ) \right] \\
   \Gamma^{\alpha\beta\mu}_{\mathrm{Z}\gamma\gamma}\rightarrow & ie~\frac{h_1^\gamma}{m_{\mathrm{Z}}^2} & \left[
 q^2_2~(q_3^\beta g^{\mu\alpha} - q_3^\alpha g^{\beta\mu})
+q^2_3~(q_2^\mu g^{\alpha\beta} - q_2^\alpha g^{\beta\mu})
                                  \right] \label{eq:lastoffshell}
\end{eqnarray}

\noindent
where the introduction of the $h_1^V,h_3^V,f_4^V,f_5^V$ parameters will 
be justified later.
The (always outgoing) four-momenta $q_j (j=1,3)$ refer to the particles
appearing in the position $j$ of the $V_1 V_2 V_3$ label. The 
following index correspondence is assumed: $1\leftrightarrow\alpha$, 
$2\leftrightarrow\beta$, $3\leftrightarrow\mu$. Terms
proportional to $q_1^\alpha, q_2^\beta$ and $q_3^\mu$ are neglected.

  When particles 1 and 2 are assumed to be on-shell bosons, 
the previous expressions become:
\begin{eqnarray}
  \Gamma^{\alpha\beta\mu}_{\mathrm{Z}\mathrm{Z}\mathrm{Z}}\rightarrow & ie~\frac{f_5^Z}{m_{\mathrm{Z}}^2} & \left[ 
   (q^2_V-m_{\mathrm{Z}}^2)~
   \epsilon^{\alpha\beta\mu\rho}~(q_{1\rho}-q_{2\rho}) \right] 
        \label{eq:firstonshell} \\
  \Gamma^{\alpha\beta\mu}_{\mathrm{Z}\gamma\mathrm{Z}}\rightarrow & ie~\frac{h_3^Z}{m_{\mathrm{Z}}^2} & \left[
    (q^2_V-m_{\mathrm{Z}}^2)~\epsilon^{\alpha\beta\mu\rho}~q_{2\rho} \right] \\
  \Gamma^{\alpha\beta\mu}_{\mathrm{Z}\mathrm{Z}\gamma}\rightarrow & ie~\frac{f_5^\gamma}{m_{\mathrm{Z}}^2} & \left[
  q^2_V~\epsilon^{\alpha\beta\mu\rho}~(q_{1\rho}-q_{2\rho}) \right] \\
  \Gamma^{\alpha\beta\mu}_{\mathrm{Z}\gamma\gamma}\rightarrow & ie~\frac{h_3^\gamma}{m_{\mathrm{Z}}^2} & \left[
    q^2_V~\epsilon^{\alpha\beta\mu\rho}~q_{2\rho} \right] \\
   \Gamma^{\alpha\beta\mu}_{\mathrm{Z}\mathrm{Z}\mathrm{Z}}\rightarrow & ie~\frac{f_4^Z}{m_{\mathrm{Z}}^2} & \left[
     (q^2_V-m_{\mathrm{Z}}^2)~(
   g^{\beta\mu}q_V^\alpha +
   g^{\mu\alpha}q_V^\beta ) \right] \\
   \Gamma^{\alpha\beta\mu}_{\mathrm{Z}\gamma\mathrm{Z}}\rightarrow & ie~\frac{h_1^Z}{m_{\mathrm{Z}}^2} & \left[
   (q^2_V-m_{\mathrm{Z}}^2)~(g^{\alpha\beta} q_2^\mu - g^{\beta\mu} q_2^\alpha) 
      \right]\\
   \Gamma^{\alpha\beta\mu}_{\mathrm{Z}\mathrm{Z}\gamma}\rightarrow & ie~\frac{f_4^\gamma}{m_{\mathrm{Z}}^2} & \left[ 
    q^2_V~(
   g^{\beta\mu}q_V^\alpha + g^{\mu\alpha}q_V^\beta ) \right] \\
   \Gamma^{\alpha\beta\mu}_{\mathrm{Z}\gamma\gamma}\rightarrow & ie~\frac{h_1^\gamma}{m_{\mathrm{Z}}^2} & \left[
   q^2_V~(g^{\alpha\beta} q_2^\mu - g^{\beta\mu} q_2^\alpha) \right]
                                    \label{eq:lastonshell}
\end{eqnarray}

\noindent
where $q_V\equiv -q_3$ (ingoing four-momentum).

  No new terms are found when the final on-shell particles are assumed to 
be 1 and 3, or 2 and 3~\footnote{The surviving terms differ by trivial 
interchanges of identical bosons indices.}.
 Structures \ref{eq:firstonshell}-\ref{eq:lastonshell} exhaust all the 
on-shell possibilities among neutral gauge bosons. Now the reason for 
introducing $h_1^V,h_3^V,f_4^V$ 
and $f_5^V$ becomes evident: all terms lead to the usual convention of 
equations \ref{eq:defzg}-\ref{eq:defzz} in the on-shell limit. 
 This feature was also noticed in \cite{gounaris_offshell,rakshit}.

 As commented before, $h_2^V$ and $h_4^V$ couplings do not appear here
because they are associated to Lagrangians of higher dimension.
Concerning the most general off-shell vertex structures
\ref{eq:firstoffshell}-\ref{eq:lastoffshell}, some important comments are 
necessary:

\begin{itemize}
  \item[a)] The introduction of the $h_j^V$ and $f_j^V$ couplings in this 
context implies
a redefinition of the convention in present $\mathrm{e^+ e^-}\rightarrow\mathrm{Z}\mathrm{Z}$ and $\mathrm{e^+ e^-}\rightarrow\mathrm{Z}\gamma$ analyses. 
However, the next sections will show that off-shell and on-shell expressions 
lead to similar results at the experimental level.

  \item[b)] The inclusion 
of off-shell structures is theoretically well motivated, but it
does not imply that experiments should search for anomalous effects in 
regions with dominant off-shell boson production. The maximal sensitivity 
is always provided by the analysis of 
$\mathrm{e^+ e^-}\rightarrow\mathrm{Z}^*\gamma\rightarrow\mathrm{f}\bar{\mathrm{f}}\gamma$ and 
$\mathrm{e^+ e^-}\rightarrow\mathrm{Z}^*\mathrm{Z}^*\rightarrow\mathrm{f}\bar{\mathrm{f}}\mathrm{f^\prime}\mathrm{\overline{f}^\prime}$ events in the vicinity of 
the $\mathrm{Z}$ resonances, corresponding to a sensible signal definition 
of $\mathrm{Z}\gamma$ and $\mathrm{Z}\mathrm{Z}$ final states. There, in addition,
``signal'' statistics is high and non-sensitive backgrounds are reduced.

  \item[c)] The standard $\mathrm{e^+ e^-}\rightarrow\mathrm{Z}\gamma$ and $\mathrm{e^+ e^-}\rightarrow\mathrm{Z}\mathrm{Z}$ analyses cover all reasonable 
types of vertex structures.
No additional samples are required in order to complete a search for anomalous 
effects at the lowest dimension (six). And
 these terms are guaranteed to be the ones which provide the largest 
effects from New Physics lying above the center-of-mass energy of the collision: 
$\Lambda > \sqrt{s}$.

\end{itemize}

\section{On-shell versus off-shell at the experimental level}

\indent

   Comparing equations \ref{eq:firstoffshell}-\ref{eq:lastoffshell} and
equations \ref{eq:firstonshell}-\ref{eq:lastonshell}, the following conclusions
are obtained:

\begin{itemize}
  \item The on-shell and off-shell vertex functions associated to $f_5^\gamma$ and 
$f_4^\gamma$ are identical.
  \item The on-shell and off-shell vertex functions associated to 
$h_1^\gamma$ and $h_3^\gamma$ coincide in the case of real photon production 
($q_2^2=0$), i.e. in the relevant case of $\mathrm{e^+ e^-}\rightarrow\mathrm{Z}^*\gamma$ production.
  \item The on-shell and off-shell vertex functions associated to 
$h_j^Z,f_j^Z$ differ by additive terms of order
${\displaystyle \frac{q_Z^2-m_{\mathrm{Z}}^2}{q_V^2-m_Z^2} \approx 
\frac{m_{\mathrm{Z}}\Gamma_{\mathrm{Z}}}{s-m_Z^2}}$. 
\end{itemize}

   Therefore, the only relevant differences between the two set of expressions
appear for $f_j^Z$ and $h_j^Z$. These differences are expected to be 
qualitatively small, but a quantitative statement is absolutely necessary in 
order to assess the validity of present experimental searches.

  In order to quantify the effects of an off-shell treatment
on present LEP results~\cite{aleph,delphi,l3,opal}, 
100000 $\mathrm{e^+ e^-}\rightarrow(\mathrm{Z}/\gamma)^*\gamma\rightarrow\mathrm{f}\bar{\mathrm{f}}\gamma$ and 
$\mathrm{e^+ e^-}\rightarrow(\mathrm{Z}/\gamma)^*(\mathrm{Z}/\gamma)^*\rightarrow\mathrm{f}\bar{\mathrm{f}}\mathrm{f^\prime}\mathrm{\overline{f}^\prime}$
events at a center-of-mass energy of $\sqrt{s}=200 {\rm \ Ge\kern -0.1em V}$ are generated.
The values $h_j^Z,f_j^Z = 0.25$, $0.5$, $1.0$, $2.0$ are considered. A more
realistic experimental scenario is obtained by selecting events in which the
two-fermion invariant masses, $m_{\mathrm{f}\bar{\mathrm{f}}}$, are consistent with the Z mass,
$\mid m_{\mathrm{f}\bar{\mathrm{f}}}-m_{\mathrm{Z}}\mid < 10 {\rm \ Ge\kern -0.1em V}$. In addition,
a cut on the polar angle of photons, $\mid\cos\theta_\gamma\mid<0.9$, 
is applied.

  For the $h_j^Z$ case, the study is performed by a reweighting 
procedure according to the $\mathrm{e^+ e^-}\rightarrow(\mathrm{Z}/\gamma)^*\gamma\rightarrow\mathrm{f}\bar{\mathrm{f}}\gamma$ 
anomalous matrix element, either 
under off-shell (equations \ref{eq:firstoffshell}-\ref{eq:lastoffshell}) or
under on-shell~\cite{hagiwara} assumptions. 

  A first observable sensitive to anomalous couplings is 
the total cross section. The relative differences between off-shell 
and on-shell cases are reported in Table 
\ref{tab:deltaN_zg}. These extremely small numbers are somehow expected, 
since off-shell deviations have similar sizes but different signs above and 
below the $\mathrm{Z}$ mass.

\begin{table}[htbp]
\begin{center}
\begin{tabular}{|c|c|}
\hline
 Coupling value & ${\displaystyle \frac{\Delta N}{N}}$ \\
\hline
\hline
 $h_1^Z=0.25\phantom{5}$ & $(0.9 \pm 0.2)~10^{-5}$ \\
 $h_1^Z=0.5\phantom{55}$ & $(3.1 \pm 0.5)~10^{-5}$ \\
 $h_1^Z=1.0\phantom{55}$ & $(0.8 \pm 0.1)~10^{-4}$ \\
 $h_1^Z=2.0\phantom{55}$ & $(1.4 \pm 0.2)~10^{-4}$ \\
\hline
 $h_3^Z=0.25\phantom{5}$ & $(0.3 \pm 0.2)~10^{-5}$ \\
 $h_3^Z=0.5\phantom{55}$ & $(2.1 \pm 0.5)~10^{-5}$ \\
 $h_3^Z=1.0\phantom{55}$ & $(0.7 \pm 0.1)~10^{-4}$ \\
 $h_3^Z=2.0\phantom{55}$ & $(1.3 \pm 0.2)~10^{-4}$ \\
\hline
\end{tabular}
\caption{Relative difference in the number of expected events,
$\Delta N/N$, between off-shell and on-shell analyses at $\sqrt{s}=200 {\rm \ Ge\kern -0.1em V}$. 
Different values of the $h_j^Z$ 
anomalous couplings are considered. Cuts on the 
fermion-pair invariant mass, $\mid m_{\mathrm{f}\bar{\mathrm{f}}}-m_{\mathrm{Z}}\mid < 10 {\rm \ Ge\kern -0.1em V}$, and 
on the photon polar angle, $\mid\cos\theta_\gamma\mid<0.9$, are applied.}
\label{tab:deltaN_zg}
\end{center}
\end{table}

  Even if the differences in the total rate are negligible, experiments use to 
combine cross section measurements and shape information in the full phase 
space. A powerful
way to study the effect of the differences in shape is by analyzing the 
mean values of the optimal observables of the process. In the general case
the differential cross section in the presence of an anomalous coupling $h$ 
can be expressed as follows:
\begin{eqnarray}
   \left. \frac{d^2 \sigma}{d O_1~d O_2} \right|_h=
   \left. \frac{d^2 \sigma}{d O_1~d O_2} \right|_{h=0}~
              \left( 1+ h~O_1 + h^2~O_2 \right)
\end{eqnarray}

\noindent
where the variables $O_1$ and $O_2$, also known as {\it optimal observables}, 
are functions of the phase space variables of the event, with no explicit 
dependence on $h$. 
 The previous equation guarantees that the maximal information on 
$h$ is obtained by a study of the event density as a function of the variables
$O_1$ and $O_2$. 

  For small CP-conserving couplings, $h_3^Z\rightarrow 0$, only 
the $O_1$ variable contributes. In fact, in the limit of vanishing couplings 
the maximal sensitivity 
is obtained by a simultaneous measurement of the total cross section and 
of the mean value of $O_1$.
For CP-violating couplings like $h_1^Z$, $O_1$ is not the relevant
variable, since CP-violating and CP-conserving (SM) terms 
do not interfere~\footnote{ This is strictly true at the same order of 
perturbative expansion. In practice, some 
interference remains due to the presence of $im_{\mathrm{Z}}\Gamma_{\mathrm{Z}}$ terms in the amplitudes, 
originating from higher order terms.}. In this case,
$O_2$ will be considered as the sensitive quantity.

\begin{figure}[htbp]
\begin{center}
    \includegraphics*[width=0.49\textwidth]{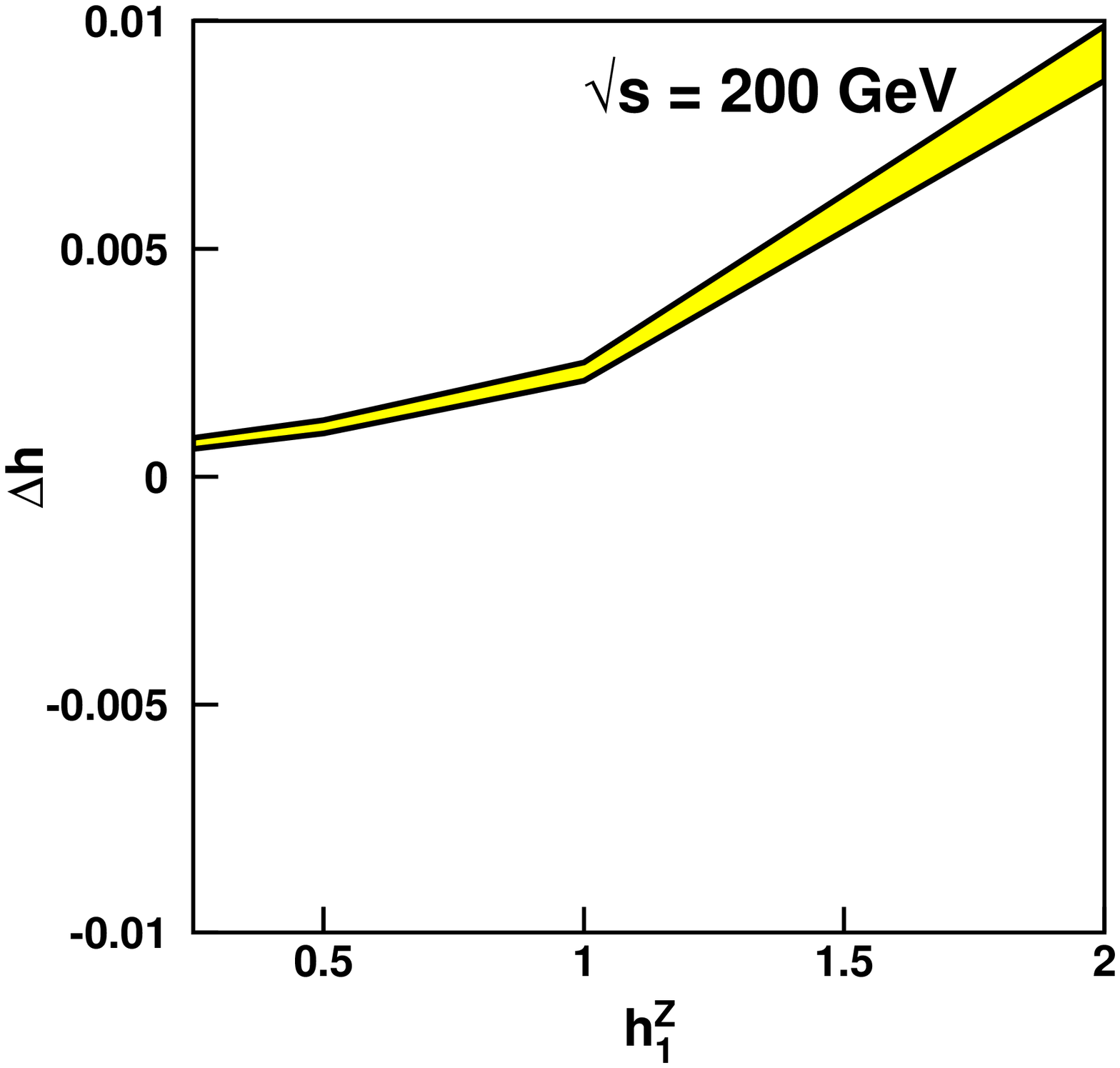}
    \includegraphics*[width=0.49\textwidth]{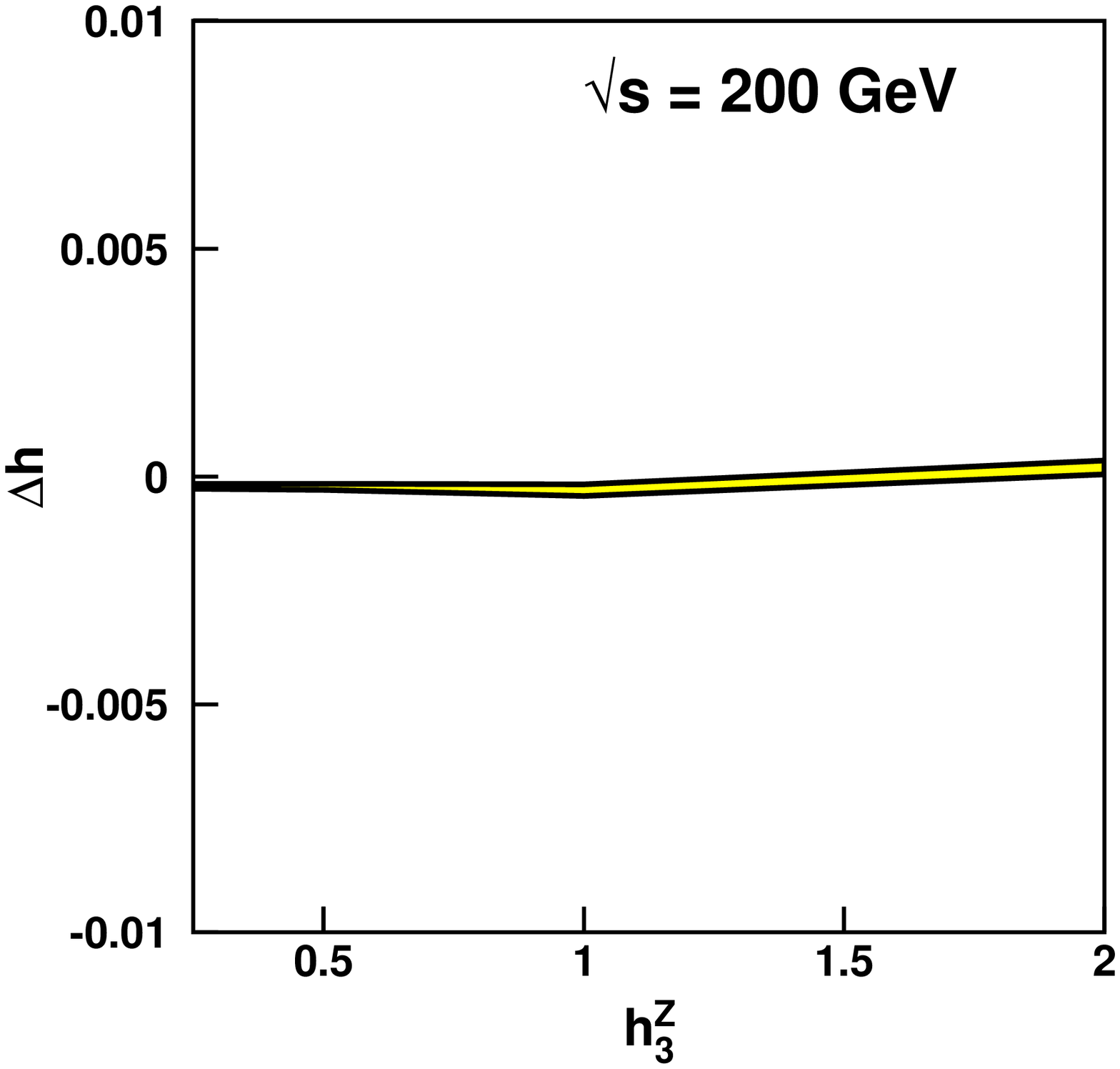}
\end{center}
\caption{ Differences, $\Delta h$, between off-shell and on-shell 
measurements of the anomalous gauge couplings $h_1^Z$ (left) 
and $h_3^Z$ (right). Measurements are derived from the mean values 
of the $O_2$ distribution (for $h_1^Z$) and of the $O_1$ distribution 
(for $h_3^Z$). The analyzed process is
$\mathrm{e^+ e^-}\rightarrow(\mathrm{Z}/\gamma)^*\gamma\rightarrow\mathrm{f}\bar{\mathrm{f}}\gamma$ at $\sqrt{s} = 200 {\rm \ Ge\kern -0.1em V}$.}
\label{fig:dh200}
\end{figure}

  Using the mean values of $O_1$ and $O_2$ as inputs, the values for the 
different couplings
are extracted. The difference observed between the measurements of a 
coupling $h$ using off-shell and on-shell approaches will be denoted by 
$\Delta h$.
It quantifies the influence of discrepancies in the shapes of phase space 
distributions between the two treatments. As observed in Figure 
\ref{fig:dh200}, the absolute differences at $\sqrt{s}=200 {\rm \ Ge\kern -0.1em V}$ never exceed 
$\mid\Delta h\mid = 0.01$ in the range under study, and are negligible  when 
compared to the present experimental uncertainties \cite{lep}.

\begin{figure}[htbp]
\begin{center}
    \includegraphics*[width=0.49\textwidth]{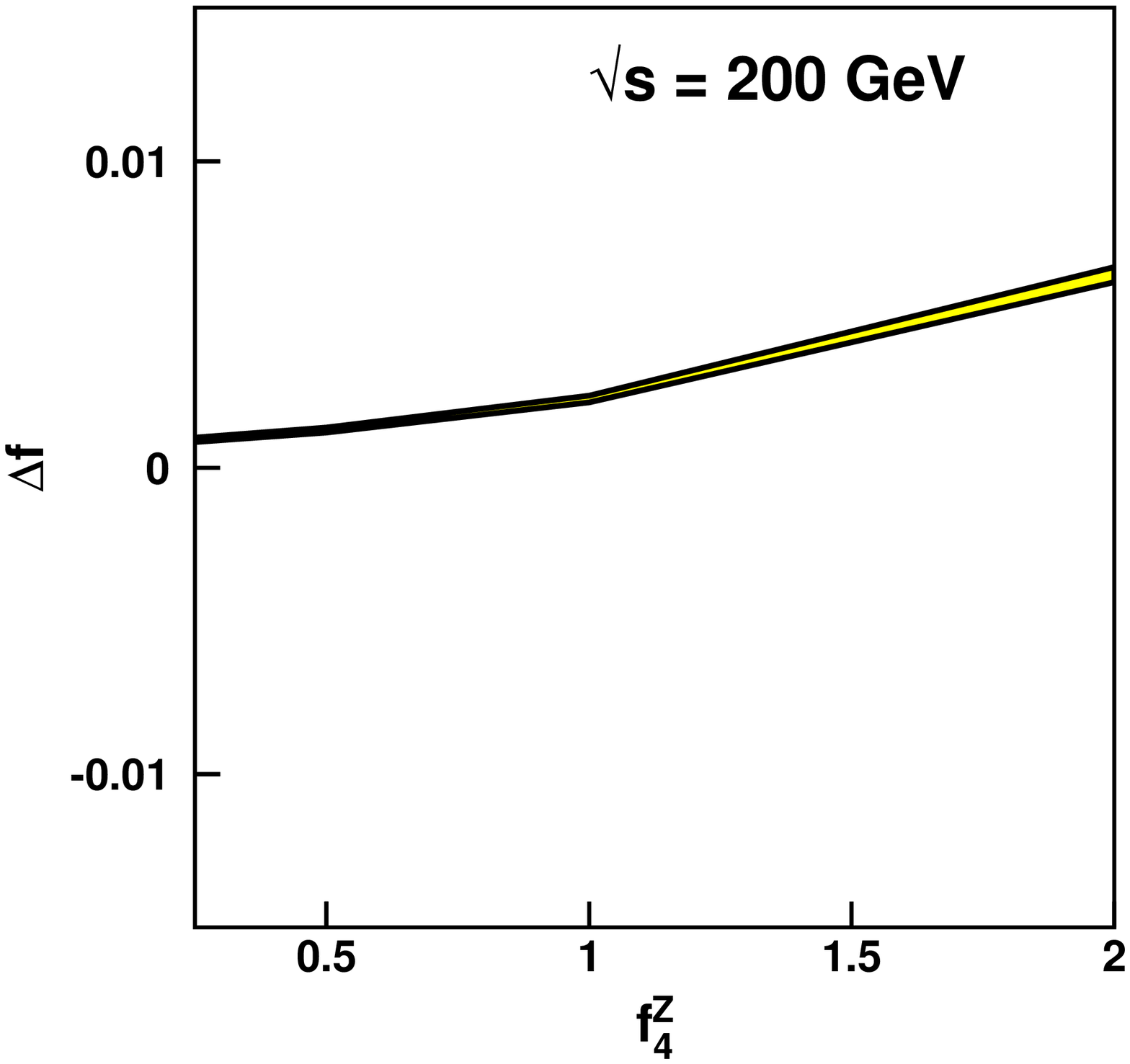}
    \includegraphics*[width=0.49\textwidth]{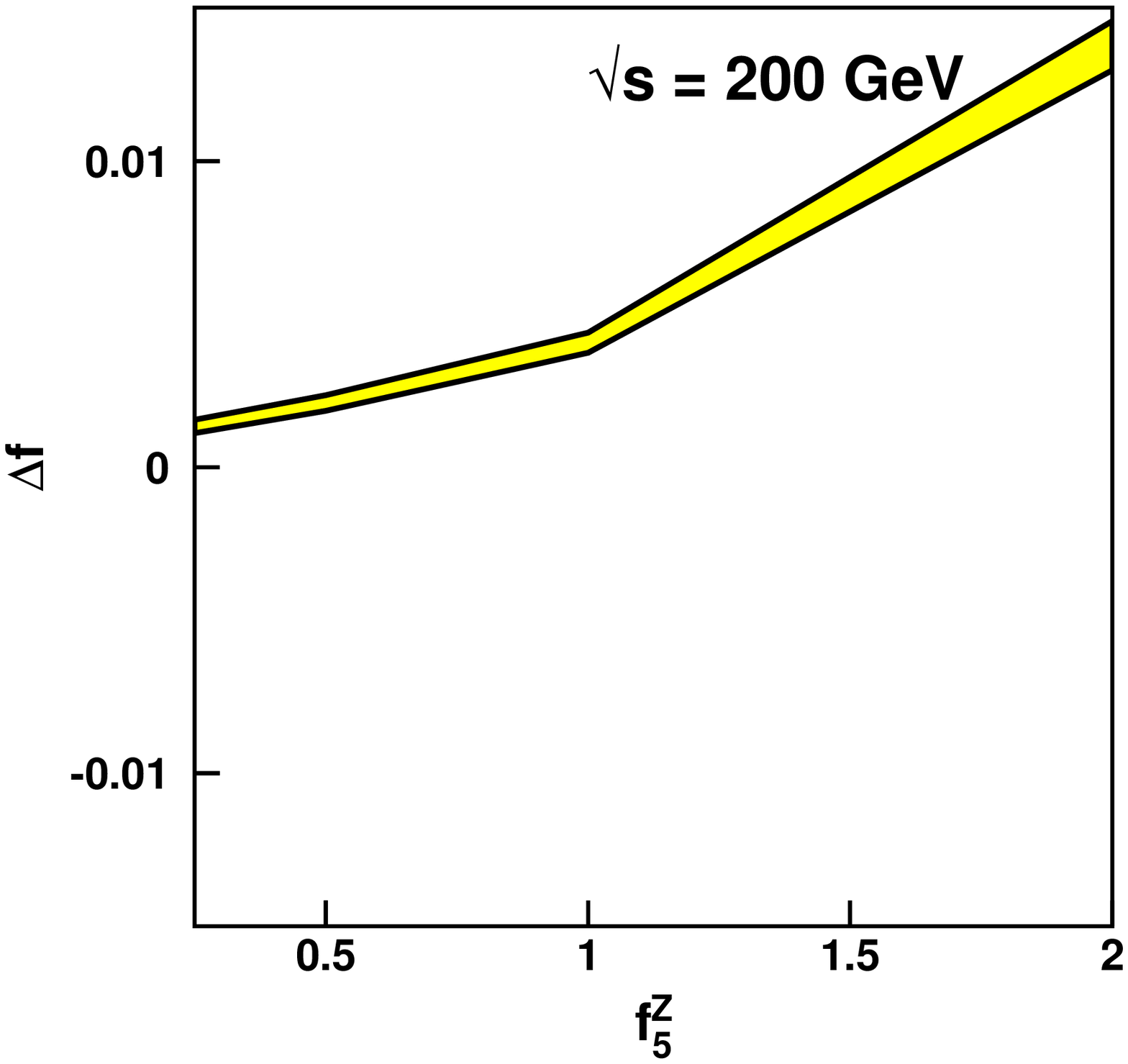}
\end{center}
\caption{ Differences, $\Delta f$, between off-shell and on-shell
measurements of the anomalous gauge couplings $f_4^Z$ (left)
and $f_5^Z$ (right). Measurements are derived from the mean values
of the $O_2$ distribution (for $f_4^Z$) and of the $O_1$ distribution
(for $f_5^Z$). The analyzed process is
$\mathrm{e^+ e^-}\rightarrow(\mathrm{Z}/\gamma)^*(\mathrm{Z}/\gamma)^*\rightarrow\mathrm{f}\bar{\mathrm{f}}\mathrm{f^\prime}\mathrm{\overline{f}^\prime}$ 
at $\sqrt{s} = 200 {\rm \ Ge\kern -0.1em V}$.}
\label{fig:df200}
\end{figure}

  For the $f_j^Z$ case, the study is performed by a reweighting 
procedure according to the 
$\mathrm{e^+ e^-}\rightarrow(\mathrm{Z}/\gamma)^*(\mathrm{Z}/\gamma)^*\rightarrow\mathrm{f}\bar{\mathrm{f}}\mathrm{f^\prime}\mathrm{\overline{f}^\prime}$ 
anomalous matrix element, either 
under off-shell (equations \ref{eq:firstoffshell}-\ref{eq:lastoffshell}) 
or on-shell~\cite{ourpaper} assumptions.
Again, the relative differences in cross section between the two approaches 
are extremely small (Table \ref{tab:deltaN_zz}).
  Similarly to the $h_j^Z$ case, the mean values of the optimal observables
give access to the values of the $f_j^Z$ couplings.
The differences between off-shell and on-shell treatments 
due to discrepancies in the shape of the phase space distributions are 
denoted by $\Delta f$. Figure \ref{fig:df200} shows that the 
differences never exceed $\mid\Delta f\mid = 0.015$, and are negligible when 
compared to the present experimental uncertainties~\cite{lep}.

\begin{table}[htbp]
\begin{center}
\begin{tabular}{|c|c|}
\hline
 Coupling value & ${\displaystyle \frac{\Delta N}{N}}$ \\
\hline
\hline
 $f_4^Z=0.25\phantom{5}$ & $(0.6 \pm 0.1)~10^{-5}$ \\
 $f_4^Z=0.5\phantom{55}$ & $(2.1 \pm 0.2)~10^{-5}$ \\
 $f_4^Z=1.0\phantom{55}$ & $(5.8 \pm 0.5)~10^{-5}$ \\
 $f_4^Z=2.0\phantom{55}$ & $(1.1 \pm 0.1)~10^{-4}$ \\
\hline
 $f_5^Z=0.25\phantom{5}$ & $(5.4 \pm 0.9)~10^{-5}$ \\
 $f_5^Z=0.5\phantom{55}$ & $(1.7 \pm 0.2)~10^{-4}$ \\
 $f_5^Z=1.0\phantom{55}$ & $(5.3 \pm 0.3)~10^{-4}$ \\
 $f_5^Z=2.0\phantom{55}$ & $(1.5 \pm 0.1)~10^{-3}$ \\
\hline
\end{tabular}
\caption{Relative difference in the number of expected events, 
$\Delta N/N$, between off-shell and on-shell analyses at $\sqrt{s}=200 {\rm \ Ge\kern -0.1em V}$.
Different values of the $f_j^Z$ 
anomalous couplings are considered. A cut on the 
relevant fermion-pair invariant masses, $\mid m_{\mathrm{f}\bar{\mathrm{f}}}-m_{\mathrm{Z}}\mid < 10 {\rm \ Ge\kern -0.1em V}$, 
is applied.}
\label{tab:deltaN_zz}
\end{center}
\end{table}

\begin{figure}[htbp]
\begin{center}
    \includegraphics*[width=0.49\textwidth]{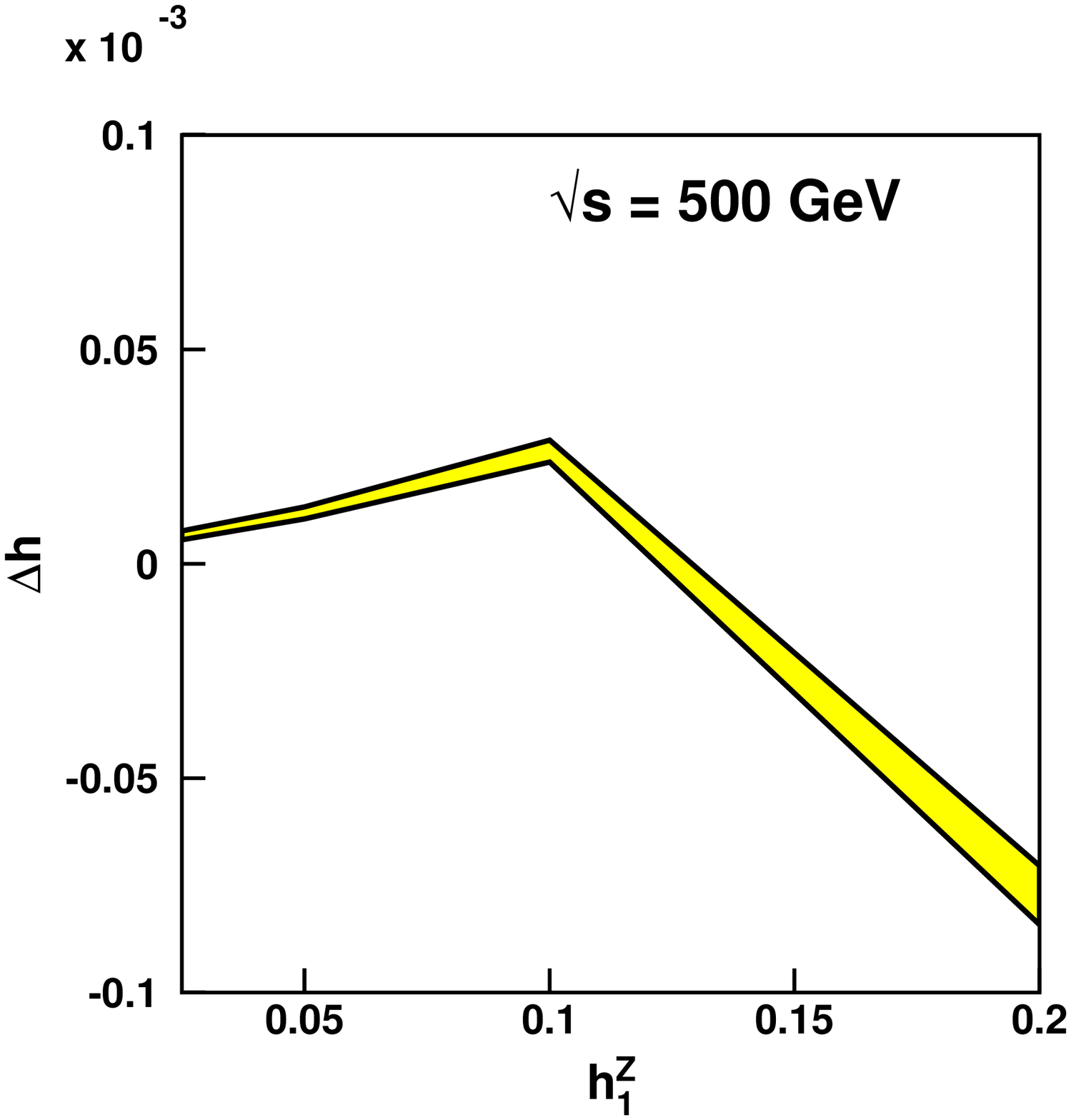}
    \includegraphics*[width=0.49\textwidth]{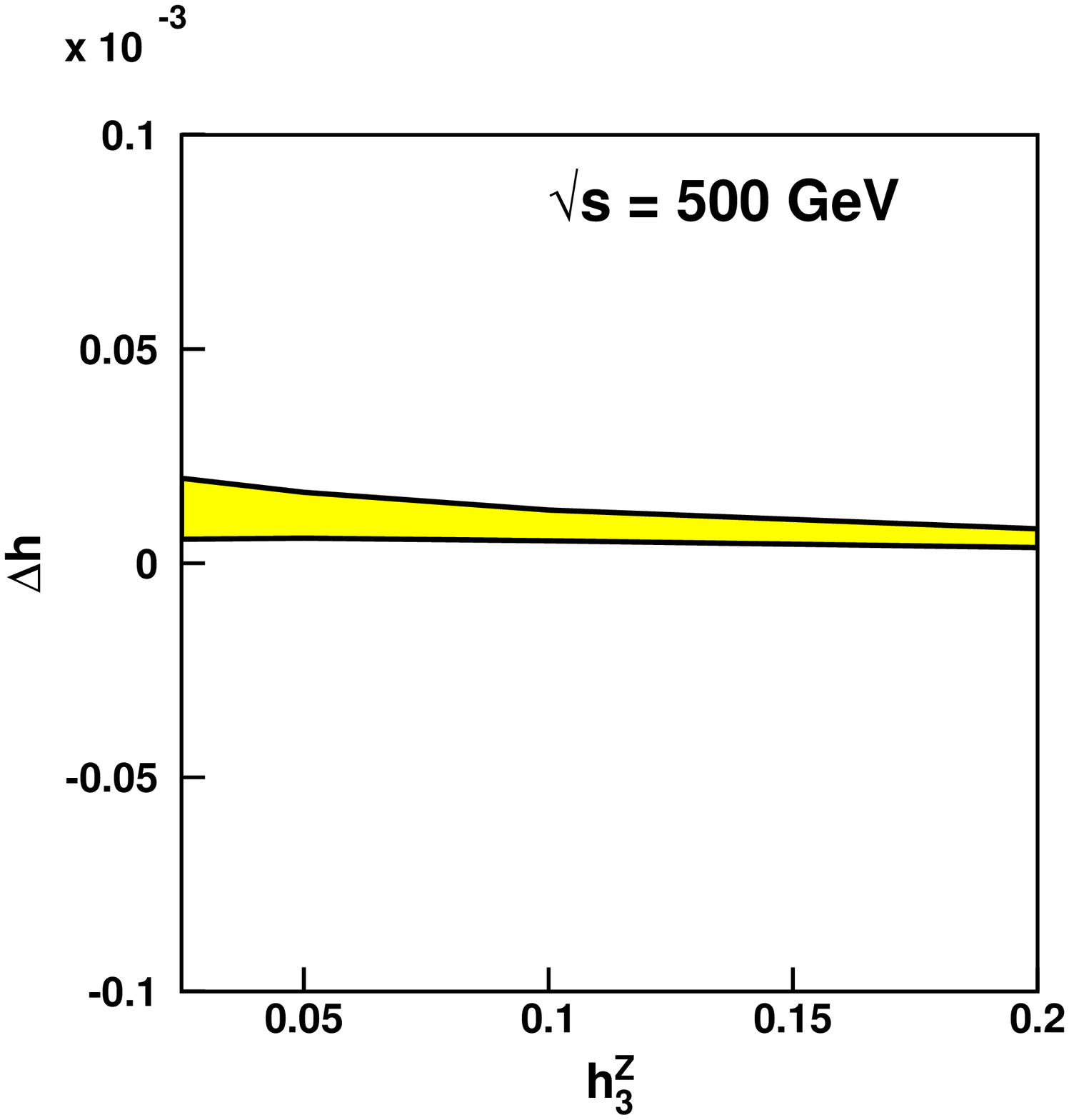}
\end{center}
\caption{ Differences, $\Delta h$, between off-shell and on-shell 
measurements of the anomalous gauge couplings $h_1^Z$ (left) 
and $h_3^Z$ (right). Measurements are derived from the mean values 
of the $O_2$ distribution (for $h_1^Z$) and of the $O_1$ distribution 
(for $h_3^Z$). The analyzed process is
$\mathrm{e^+ e^-}\rightarrow(\mathrm{Z}/\gamma)^*\gamma\rightarrow\mathrm{f}\bar{\mathrm{f}}\gamma$ at $\sqrt{s} = 500 {\rm \ Ge\kern -0.1em V}$.}
\label{fig:dh500}
\end{figure}

\begin{figure}[htbp]
\begin{center}
    \includegraphics*[width=0.49\textwidth]{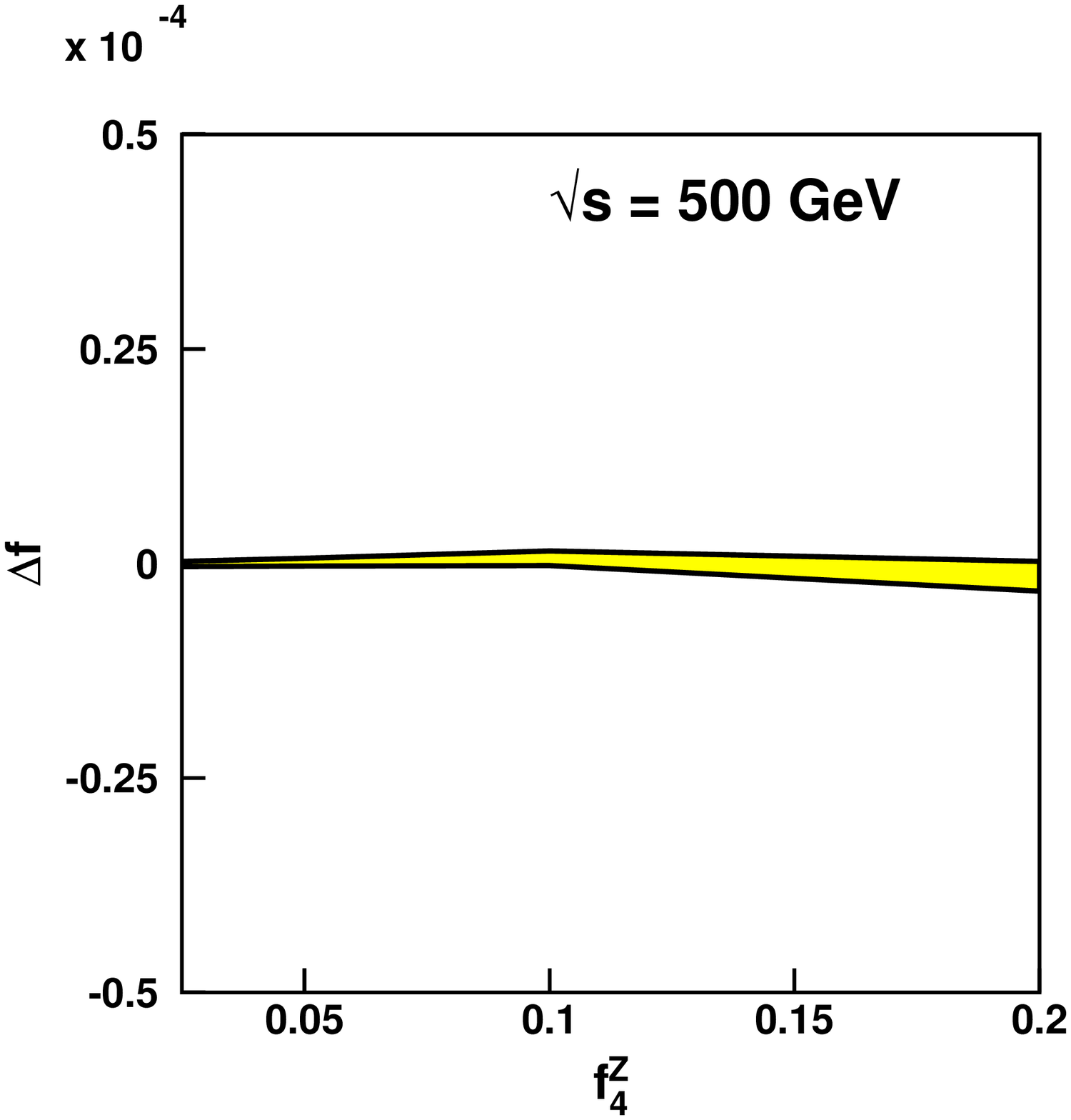}
    \includegraphics*[width=0.49\textwidth]{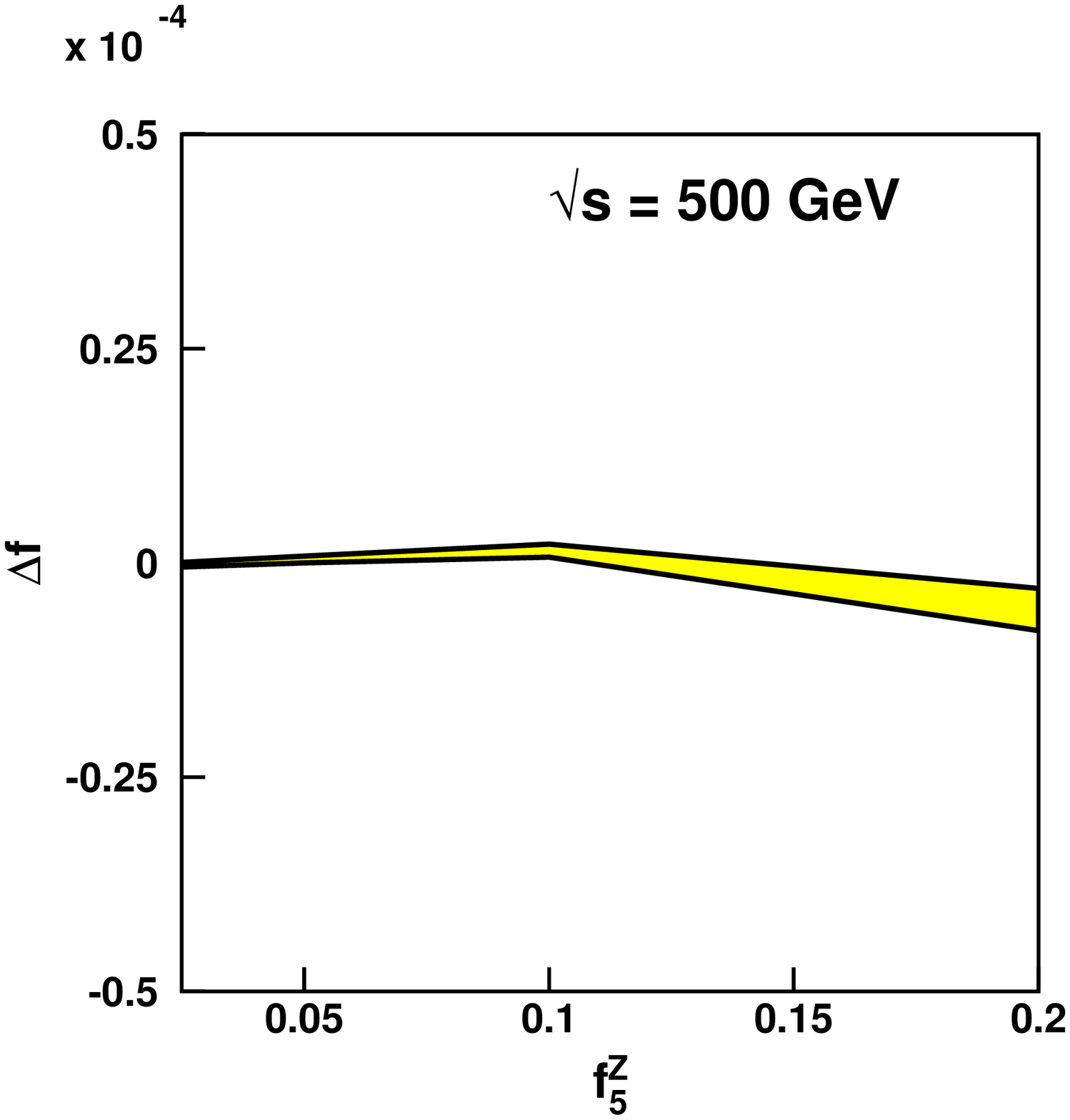}
\end{center}
\caption{ Differences, $\Delta f$, between off-shell and on-shell
measurements of the anomalous gauge couplings $f_4^Z$ (left)
and $f_5^Z$ (right). Measurements are derived from the mean values
of the $O_2$ distribution (for $f_4^Z$) and of the $O_1$ distribution
(for $f_5^Z$). The analyzed process is
$\mathrm{e^+ e^-}\rightarrow(\mathrm{Z}/\gamma)^*(\mathrm{Z}/\gamma)^*\rightarrow\mathrm{f}\bar{\mathrm{f}}\mathrm{f^\prime}\mathrm{\overline{f}^\prime}$ 
at $\sqrt{s} = 500 {\rm \ Ge\kern -0.1em V}$.}
\label{fig:df500}
\end{figure}

   In order to investigate the implications for the next 
generation of linear colliders, all previous exercises are
repeated for a center-of-mass energy of $\sqrt{s}=500 {\rm \ Ge\kern -0.1em V}$. Since the 
sensitivity at these energies is expected to be at least one order of 
magnitude larger than at $\sqrt{s}=200 {\rm \ Ge\kern -0.1em V}$ \cite{ourpaper}, 
the values $h_j^Z,f_j^Z = 0.025, 0.05, 0.1, 0.2$ are considered. Cross section
differences are shown in Table \ref{tab:deltaN500}, and the shifts
due to shape distribution discrepancies
are presented in Figures \ref{fig:dh500}-\ref{fig:df500}. It is evident that 
the differences between off-shell and on-shell treatments are extremely small
in all cases.

\begin{figure}[htbp]
\begin{center}
    \includegraphics*[width=0.49\textwidth]{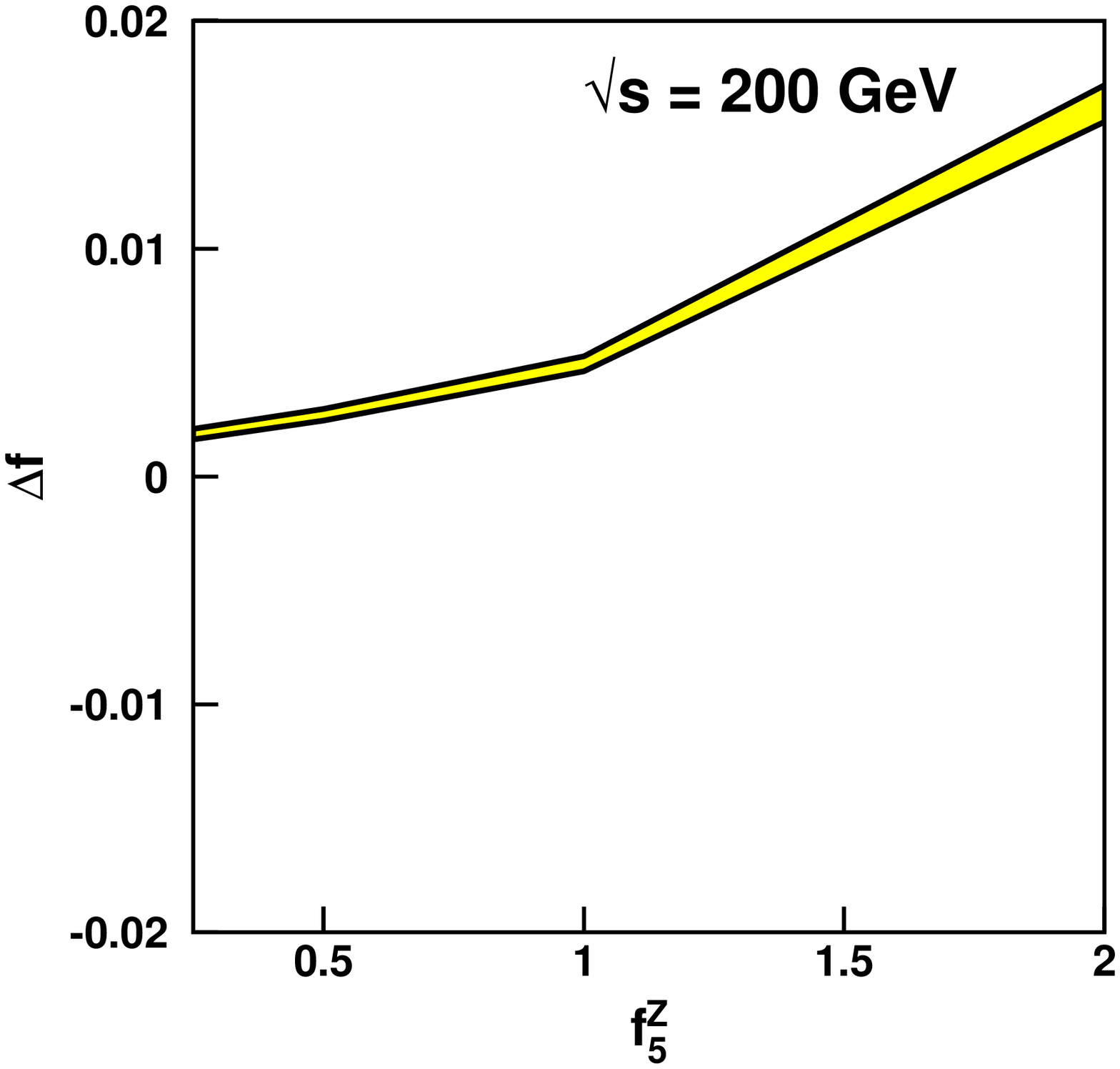}
    \includegraphics*[width=0.49\textwidth]{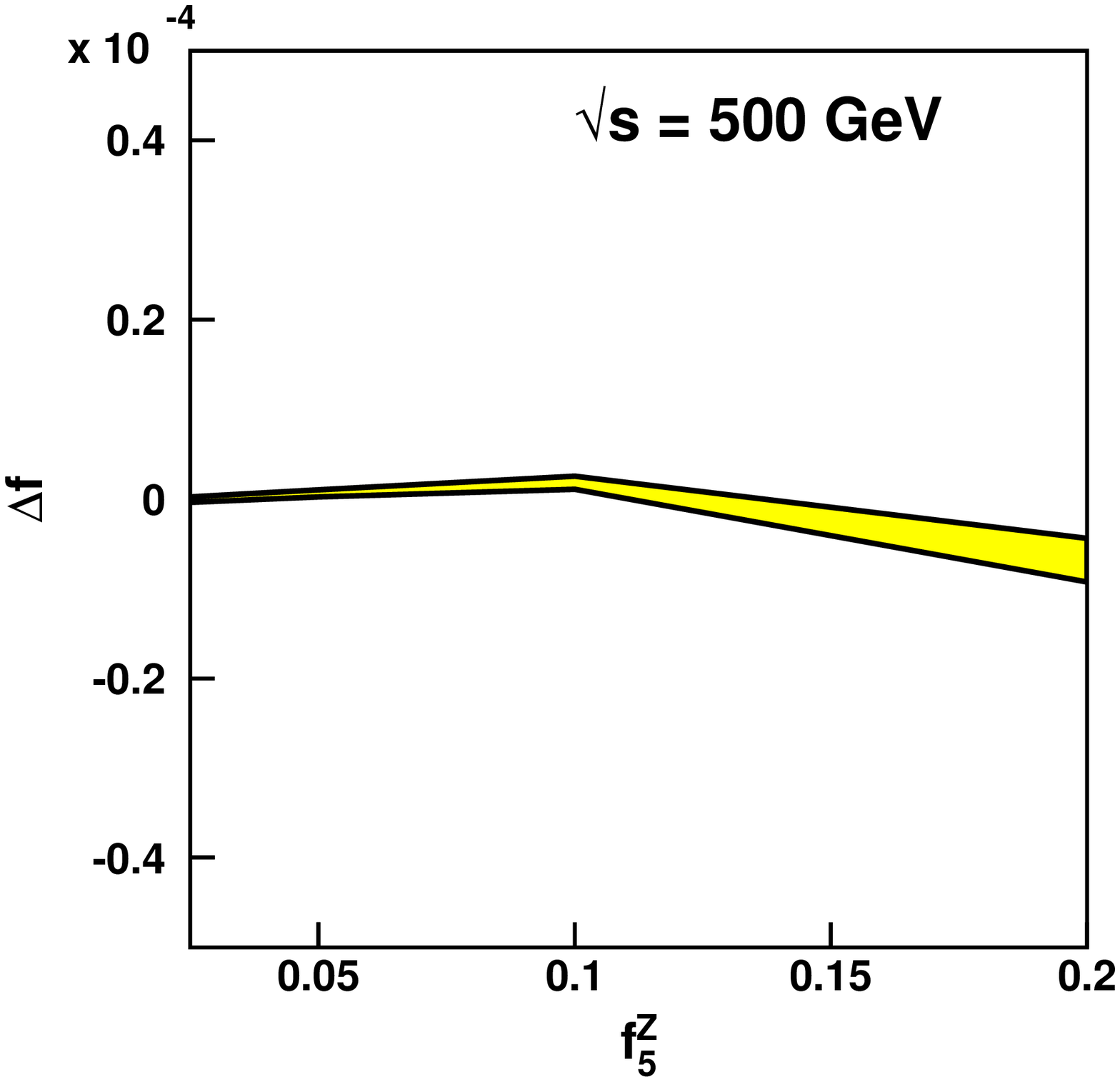}
\end{center}
\caption{ Differences, $\Delta f$, between off-shell and on-shell
measurements of the anomalous gauge coupling $f_5^Z$ at 
$\sqrt{s}=200{\rm \ Ge\kern -0.1em V}$ and 
$\sqrt{s}=500{\rm \ Ge\kern -0.1em V}$. Measurements are
derived from the mean value of the $O_1$ distribution.
The analyzed process is
$\mathrm{e^+ e^-}\rightarrow(\mathrm{Z}/\gamma)^*(\mathrm{Z}/\gamma)^*\rightarrow\mathrm{f}\bar{\mathrm{f}}\mathrm{f^\prime}\mathrm{\overline{f}^\prime}$.
A loose invariant mass cut,
$\mid m_{\mathrm{f}\bar{\mathrm{f}}}-m_{\mathrm{Z}}\mid < 50 {\rm \ Ge\kern -0.1em V}$,
is used.}
\label{fig:f5z_50}
\end{figure}

  Finally, we should investigate the effect of changing 
the mass window cut around the Z mass. This concerns the hypothetical 
case of a LEP analysis with relaxed constraints and also
the $h_j^V$ limits obtained at Tevatron \cite{cdf,d0}. At 
$\mathrm{p}\overline{\mathrm{p}}$ colliders the 
requirements of consistency with the $\mathrm{Z}$ mass are either
loose (CDF) or somehow indirect (D0 and $\mathrm{Z}\rightarrow\nu\bar{\nu}$). 
We have estimated the $h_j^Z$ and $f_j^Z$ differences between 
on-shell and off-shell approaches for an invariant mass cut of 
$\mid m_{\mathrm{f}\bar{\mathrm{f}}}-m_{\mathrm{Z}}\mid < 50 {\rm \ Ge\kern -0.1em V}$. 
The results do not differ significantly from those obtained for
$\mid m_{\mathrm{f}\bar{\mathrm{f}}}-m_{\mathrm{Z}}\mid < 10 {\rm \ Ge\kern -0.1em V}$. 
Table \ref{tab:f5z_50} and Figure \ref{fig:f5z_50} show them for
the coupling where the largest effect is        
found ($f_5^Z$). We conclude that the inclusion of 
final fermion pairs away from the Z resonance region leads to marginal 
biases in the analysis.

\begin{table}[htbp]
\begin{center}
\begin{tabular}{|c|c|}
\hline
 Coupling value & ${\displaystyle \frac{\Delta N}{N}}$ \\
\hline
\hline
 $h_1^Z=0.025$ & $\phantom{-}(0.4 \pm 0.1)~10^{-5}$ \\
 $h_1^Z=0.05\phantom{5}$ & $\phantom{-}(0.9 \pm 0.2)~10^{-5}$ \\
 $h_1^Z=0.1\phantom{55}$ & $\phantom{-}(1.5 \pm 0.3)~10^{-5}$ \\
 $h_1^Z=0.2\phantom{55}$ & $\phantom{-}(1.8 \pm 0.3)~10^{-5}$ \\
\hline
 $h_3^Z=0.025$ & $\phantom{-}(0.3 \pm 0.1)~10^{-5}$ \\
 $h_3^Z=0.05\phantom{5}$ & $\phantom{-}(0.9 \pm 0.2)~10^{-5}$ \\
 $h_3^Z=0.1\phantom{55}$ & $\phantom{-}(1.4 \pm 0.3)~10^{-5}$ \\
 $h_3^Z=0.2\phantom{55}$ & $\phantom{-}(1.7 \pm 0.3)~10^{-5}$ \\
\hline
 $f_4^Z=0.025$ & $-(0.8 \pm 0.9)~10^{-6}$ \\
 $f_4^Z=0.05\phantom{5}$ & $-(1.6 \pm 1.6)~10^{-6}$ \\
 $f_4^Z=0.1\phantom{55}$ & $-(2.1 \pm 2.0)~10^{-6}$ \\
 $f_4^Z=0.2\phantom{55}$ & $-(2.3 \pm 2.2)~10^{-6}$ \\
\hline
 $f_5^Z=0.025$ & $\phantom{-}(0.7 \pm 0.1)~10^{-5}$ \\
 $f_5^Z=0.05\phantom{5}$ & $\phantom{-}(1.5 \pm 0.2)~10^{-5}$ \\
 $f_5^Z=0.1\phantom{55}$ & $\phantom{-}(1.9 \pm 0.3)~10^{-5}$ \\
 $f_5^Z=0.2\phantom{55}$ & $\phantom{-}(2.0 \pm 0.3)~10^{-5}$ \\
\hline
\end{tabular}
\caption{Relative difference in the number of expected events, 
$\Delta N/N$, between off-shell and on-shell analyses at $\sqrt{s}=500 {\rm \ Ge\kern -0.1em V}$. 
Different values of the $h_j^Z$ and $f_j^Z$ 
anomalous couplings are considered. Cuts on the 
fermion-pair invariant mass, $\mid m_{\mathrm{f}\bar{\mathrm{f}}}-m_{\mathrm{Z}}\mid < 10 {\rm \ Ge\kern -0.1em V}$, and 
on the photon polar angle, $\mid\cos\theta_\gamma\mid<0.9$, are applied.}
\label{tab:deltaN500}
\end{center}
\end{table}

\begin{table}[htbp]
\begin{center}
\begin{tabular}{|c|c|c|}
\hline
 Coupling value & $\sqrt{s}$ $({\rm Ge\kern -0.1em V})$& ${\displaystyle \frac{\Delta N}{N}}$ \\
\hline
\hline
 $f_5^Z=0.25\phantom{5}$ & $200$ & $\phantom{-}(7.1 \pm 0.9)~10^{-5}$ \\
 $f_5^Z=0.5\phantom{55}$ & $200$ & $\phantom{-}(2.0 \pm 0.2)~10^{-4}$ \\
 $f_5^Z=1.0\phantom{55}$ & $200$ & $\phantom{-}(6.0 \pm 0.4)~10^{-4}$ \\
 $f_5^Z=2.0\phantom{55}$ & $200$ & $\phantom{-}(1.6 \pm 0.1)~10^{-5}$ \\
\hline
 $f_5^Z=0.025$ & $500$ & $\phantom{-}(0.8 \pm 0.1)~10^{-5}$ \\
 $f_5^Z=0.05\phantom{5}$ & $500$ & $\phantom{-}(1.6 \pm 0.2)~10^{-5}$ \\
 $f_5^Z=0.1\phantom{55}$ & $500$ & $\phantom{-}(2.0 \pm 0.3)~10^{-5}$ \\
 $f_5^Z=0.2\phantom{55}$ & $500$ & $\phantom{-}(2.1 \pm 0.3)~10^{-5}$ \\
\hline
\end{tabular}
\caption{Relative difference in the number of expected events, 
$\Delta N/N$, between off-shell and on-shell analyses at 
$\sqrt{s}=200,~500 {\rm \ Ge\kern -0.1em V}$. 
Different values of the $f_5^Z$ 
anomalous coupling are considered. The cut on the 
fermion-pair invariant masses is
$\mid m_{\mathrm{f}\bar{\mathrm{f}}}-m_{\mathrm{Z}}\mid < 50 {\rm \ Ge\kern -0.1em V}$.}
\label{tab:f5z_50}
\end{center}
\end{table}

   So far, we have only considered the case in which $h_j^V$ and 
$f_j^V$ couplings are studied separately in 
$\mathrm{e^+ e^-}\rightarrow(\mathrm{Z}/\gamma)^*\gamma$ 
and $\mathrm{e^+ e^-}\rightarrow(\mathrm{Z}/\gamma)^*(\mathrm{Z}/\gamma)^*$ 
events. However, once off-shell effects are included, combined searches 
may become a complicated issue. An example is 
the search for $f_j^V$ couplings in the
$\mathrm{e^+ e^-}\rightarrow(\mathrm{Z}/\gamma)^*(\mathrm{Z}/\gamma)^*$ sample. In this case deviations 
due to the simultaneous presence of $h_j^V$ couplings 
(affecting the non-resonant $\mathrm{e^+ e^-}\rightarrow\mathrm{Z}^*\gamma^*$ component) may arise.

 Although tiny effects are expected in regions in which the 
the ${\displaystyle \left| 
\frac{M(\mathrm{e^+ e^-}\rightarrow\mathrm{Z}^*\gamma^*)}{M(\mathrm{e^+ e^-}\rightarrow\mathrm{Z}^*\mathrm{Z}^*)}\right|}$ 
matrix element ratio is small, 
a full off-shell treatment is advisable in general.

\section{An alternative view. Redefinition of the 
${\mathbf h_2^V}$ and ${\mathbf h_4^V}$ convention}

\indent

    Actually, the problem with the convention in Equations 
\ref{eq:defzg}-\ref{eq:defzz} can be solved at 
the ``construction'' level, just by imposing Bose-Einstein symmetry 
and electromagnetic gauge invariance as constraints.

 Let us first consider the $f_5^Z$ case.
What is relevant in the definition is the basic 
P-violating structure $i~\epsilon^{\alpha\beta\mu\rho}~q_{1\rho}$. On it 
we have to impose Bose-Einstein symmetry for the three $\mathrm{Z}$ bosons. 
It can be seen that a symmetrization of 
$i~\epsilon^{\alpha\beta\mu\rho}~(q_{1\rho}-q_{2\rho})$
leads to a trivial vanishing result. Therefore, one has to 
multiply it by a momentum-dependent scalar factor (corresponding 
to a higher dimension term): 
\begin{eqnarray}
ie~f_5^Z~\epsilon^{\alpha\beta\mu\rho}~(q_{1\rho}-q_{2\rho}) & \rightarrow & 
ie~f_5^Z~\frac{q_3^2}{m_{\mathrm{Z}}^2}~\epsilon^{\alpha\beta\mu\rho}~
(q_{1\rho}-q_{2\rho})
\end{eqnarray}

  It is the symmetrization of this last expression~\footnote{
Scalar factors like $(q_1^2+q_2^2)$ and $(q_1q_2)$ lead to an equivalent 
result.} which leads to the off-shell equation \ref{eq:firstoffshell}.
  A second example concerns the $h_1^\gamma$ coupling. In this case, we have
to impose not only Bose-Einstein symmetry, but electromagnetic gauge 
invariance on the 
P-conserving term $i~(q_2^\mu g^{\alpha\beta}-q_2^\alpha g^{\beta\mu})$.
This last requirement 
reads: $q_{3\mu}\Gamma_{\mathrm{Z}\gamma\gamma}^{\alpha\beta\mu} = 
q_{2\beta}\Gamma_{\mathrm{Z}\gamma\gamma}^{\alpha\beta\mu} = 0$, but, since
terms proportional to $q_{2\beta}, q_{3\mu}$ are neglected, the right 
expressions 
to use are: $q_{2\beta}\Gamma_{\mathrm{Z}\gamma\gamma}^{\alpha\beta\mu} \propto q_2^2$, 
$q_{3\mu}\Gamma_{\mathrm{Z}\gamma\gamma}^{\alpha\beta\mu} \propto q_3^2$. The two constraints are
satisfied by the following modification:
\begin{eqnarray}
ie~h_1^\gamma~(q_2^\mu g^{\alpha\beta}-q_2^\alpha g^{\beta\mu}) & \rightarrow & 
ie~h_1^\gamma~\frac{q_3^2}{m_{\mathrm{Z}}^2}
(q_2^\mu g^{\alpha\beta}-q_2^\alpha g^{\beta\mu})
\end{eqnarray}

  Again it is the symmetrization of this last expression
which leads to the off-shell equation \ref{eq:lastoffshell}.

  Let us now discuss the issue of anomalous couplings proceeding via 
higher dimension Lagrangians. Even if more off-shell structures, not 
covered by $h_2^V$ and $h_4^V$ on-shell studies, are possible in this case~
\cite{gounaris_offshell}, the experimental sensitivity to those new terms is
extremely low. Besides the fact that they correspond to effects from terms 
of higher dimension, they vanish exactly for $\mathrm{Z},\gamma$ on-shell 
production, whereas
a reasonable rate of off-shell boson production is required in order to 
perform a sensible measurement. Let us also remind that the most general
parametrization used in the search for 
${\mathrm W}{\mathrm W}V$ anomalous couplings 
\cite{hagiwara} neglects terms vanishing in the on-shell limit.

  Therefore, only terms associated in the on-shell limit to $h_2^V$ and 
$h_4^V$ structures will be considered. Imposing Bose-Einstein symmetry 
and electromagnetic gauge invariance the following expressions are 
obtained:
\begin{eqnarray}
\Gamma^{\alpha\beta\mu}_{\mathrm{Z}\gamma\mathrm{Z}}\rightarrow & ie~\frac{h_2^Z}{m_{\mathrm{Z}}^4} & \left[
  q_3^2~q_3^\alpha~(q_2 q_3~g^{\beta\mu}-q_2^\mu~q_3^\beta) +
  q_1^2~q_1^\mu~(q_2 q_1~g^{\alpha\beta}-q_2^\alpha~q_1^\beta) 
   \right] \nonumber \\
                                  & + ie~\frac{h_2^Z}{2m_{\mathrm{Z}}^2} & \left[
   (q_3^2-q_1^2)~(q_2^\mu g^{\alpha\beta}-q_2^\alpha g^{\beta\mu}) 
    \right] \label{eq:h2zoffshell} \\
\Gamma^{\alpha\beta\mu}_{\mathrm{Z}\gamma\gamma}\rightarrow & ie~\frac{h_2^\gamma}{m_{\mathrm{Z}}^4} & \left[
  (q_3^\alpha q_3^2+q_2^\alpha q_2^2)~
  (q_2 q_3~g^{\beta\mu}-q_2^\mu~q_3^\beta) \right] \label{eq:h2goffshell} \\
\Gamma^{\alpha\beta\mu}_{\mathrm{Z}\gamma\mathrm{Z}}\rightarrow & ie~\frac{h_4^Z}{m_{\mathrm{Z}}^4} & \left[
  q_3^2 q_3^\alpha~\epsilon^{\mu\beta\rho\sigma}~q_{3\rho}~q_{2\sigma} +
  q_1^2 q_1^\mu~\epsilon^{\alpha\beta\rho\sigma}~q_{1\rho}~q_{2\sigma} 
   \right] \nonumber \\
                                  & + ie~\frac{h_4^Z}{2m_{\mathrm{Z}}^2} & \left[
  (q_3^2-q_1^2)~\epsilon^{\alpha\beta\mu\rho}~q_{2\rho}
   \right] \label{eq:h4zoffshell} \\
\Gamma^{\alpha\beta\mu}_{\mathrm{Z}\gamma\gamma}\rightarrow & ie~\frac{h_4^\gamma}{m_{\mathrm{Z}}^4} & \left[
  (q_3^\alpha q_3^2-q_2^\alpha q_2^2)~
  \epsilon^{\mu\beta\rho\sigma}~q_{3\rho}~q_{2\sigma} \right]
    \label{eq:h4goffshell}
\end{eqnarray}

   In the on-shell limit the corresponding structures in Equation
\ref{eq:defzg} are obtained. Let us comment at this 
point that imposing Bose-Einstein symmetry on the original proposal for 
$h_2^Z$ and $h_4^Z$ couplings forces the inclusion of redundant 
structures of the $h_1^Z$ and $h_3^Z$ type, as it can be easily confirmed by 
visual inspection of Equations \ref{eq:h2zoffshell} and 
\ref{eq:h4zoffshell}. 

\section{${\mathbf SU(2)_L\times U(1)_Y}$ symmetry}

Given the good agreement between present data and
SM predictions, any signal of new physics from a 
large scale $\Lambda$ will most probably manifest at the electroweak 
scale as deviations respecting the gauge symmetry of the SM.
Concerning NTGCs, we must consider all terms containing 
neutral gauge bosons and Higgs fields in the linear realization of the
$SU(2)_L\times U(1)_Y$ gauge symmetry. Eight operators manifest at the 
lowest dimension (eight):
\begin{eqnarray}
   {\mathcal O}^A_8 & = &
    i\tilde{B}_{\mu\nu} (\partial_\sigma B^{\sigma\mu})(\Phi^+D^\nu \Phi) 
       ~+~h.c. \\
   {\mathcal O}^B_8 & = &
    i\tilde{B}_{\mu\nu} (\partial_\sigma W^{\sigma\mu}_I)
                                 (\Phi^+ \tau_I D^\nu \Phi) ~+~h.c. \\
   {\mathcal O}^C_8 & = &
    i\tilde{W}_{I\mu\nu} (\partial_\sigma B^{\sigma\mu})
                                 (\Phi^+\tau_I D^\nu \Phi) ~+~h.c. \\
   {\mathcal O}^D_8 & = &
    i\tilde{W}_{I\mu\nu} (\partial_\sigma W^{\sigma\mu}_I)
                                 (\Phi^+ D^\nu \Phi) ~+~h.c. \\
   \tilde{\mathcal O}^A_8 & = &
    i B_{\mu\nu} (\partial_\sigma B^{\sigma\mu})(\Phi^+D^\nu \Phi) 
       ~+~h.c. \\
   \tilde{\mathcal O}^B_8 & = &
    i B_{\mu\nu} (\partial_\sigma W^{\sigma\mu}_I)
                                 (\Phi^+ \tau_I D^\nu \Phi) ~+~h.c. \\
   \tilde{\mathcal O}^C_8 & = &
    i W_{I\mu\nu} (\partial_\sigma B^{\sigma\mu})
                                 (\Phi^+\tau_I D^\nu \Phi) ~+~h.c. \\
   \tilde{\mathcal O}^D_8 & = &
    i W_{I\mu\nu} (\partial_\sigma W^{\sigma\mu}_I)
                                 (\Phi^+ D^\nu \Phi) ~+~h.c.
\end{eqnarray}

\noindent
where $B^{\mu\nu}$ and $W^{\mu\nu}_I$ ($I=1,3$) are the tensor field 
associated to the $U(1)_Y$ and $SU(2)_L$ groups, respectively, 
$\tau_I$ are the Pauli matrices, 
$\Phi$ is the Higgs field and $D$ denotes the covariant derivative. The 
first four operators conserve CP and the last four operators
are CP-violating. 

  These eight operators give independent contributions to the 
$h_1^V, h_3^V, f_4^V, f_5^V$ couplings discussed in previous sections. 
Therefore, no $SU(2)_L\times U(1)_Y$ constraints among NTGC couplings 
can be imposed. Even under the extreme
assumption of fully vanishing C-violating ${\mathrm W}{\mathrm W}V$ couplings 
--$g_4^Z$, $g_5^Z$, $g_4^\gamma$, $g_5^\gamma$~\cite{hagiwara}--, constraints 
among NTGCs are weak, since these four charged couplings compete with eight 
different neutral effects. It can be shown that operators 
${\mathcal O}^A_8$ and $\tilde{\mathcal O}^A_8$ do not contain 
${\mathrm W}{\mathrm W}V$ couplings, whereas the 
operators ${\mathcal O}^B_8$ and $\tilde{\mathcal O}^B_8$ have no effect on 
${\mathrm W}{\mathrm W}V$ couplings for on-shell $W$ bosons. This last feature
follows trivially from the on-shell relation:
$\partial_\sigma W^{\sigma\mu} = -m_{\mathrm{W}}^2 W^\mu$. 
If only ${\mathcal O}^A_8$, ${\mathcal O}^B_8$, $\tilde{\mathcal O}^A_8$ 
and $\tilde{\mathcal O}^B_8$ are allowed, then the following constraints 
among NTGCs are found:
\begin{eqnarray}
   f_5^Z & = & h_3^Z\tan\theta_w \\
   f_5^\gamma & = & h_3^\gamma\tan\theta_w \\
   f_4^Z & = & h_1^Z\tan\theta_w \\
   f_4^\gamma & = & h_1^\gamma\tan\theta_w 
\end{eqnarray}

\noindent
where $\theta_w$ is the Weinberg angle.
  Our conclusions are different from those
of Reference~\cite{gounaris_offshell}, where only
operators containing {\it exclusively} neutral gauge bosons and Higgs fields
are considered as relevant and strong constraints
among NTGCs are presented.

\section{Conclusions}

\indent

   We have analyzed the experimental consequences of including a proper
off-shell treatment in the searches for anomalous NTGCs.
We find that the quantitative differences between on-shell 
and off-shell treatments are negligible, provided that the 
$\mathrm{e^+ e^-}\rightarrow\mathrm{Z}\gamma$ and $\mathrm{e^+ e^-}\rightarrow\mathrm{Z}\mathrm{Z}$ analyses are performed 
in regions where $\mathrm{Z}$ resonant production is dominant. This 
conclusion is also valid for future $\mathrm{e^+ e^-}$ studies at higher energies. 
Present on-shell studies guarantee a coverage 
of all physics deviations for which a reasonable experimental sensitivity 
is expected.
Just for theoretical consistency, and in order to avoid misleading results in 
off-resonance studies, we advocate the use of the new vertex functions
presented in Equations \ref{eq:firstoffshell}-\ref{eq:lastoffshell} and
\ref{eq:h2zoffshell}-\ref{eq:h4goffshell}. Contrary to what has been 
recently suggested in the literature~\cite{gounaris_offshell}, we find that 
only the additional assumption of vanishing C-violating charged gauge 
couplings in the ${\mathrm e}^+{\mathrm e}^-\rightarrow W^+ W^-$ process may 
lead to some $SU(2)_L\times U(1)_Y$ constraints among NTGCs.

\section*{Acknowledgements}

  We would like to thank Helge Voss and Robert Sekulin 
for interesting discussions and suggestions concerning the paper and 
the application of $SU(2)\times U(1)$ constraints at LEP.

\bibliographystyle{l3style}
\bibliography{ntgc}

\end{document}